# Topological delocalization and tuning of surface channel separation in $Bi_2Se_2Te$ Topological Insulator Thin films

Radha Krishna Gopal, Sourabh Singh, Arpita Mandal, Jit Sarkar and Chiranjib Mitra*

Indian Institute of Science Education and Research, Kolkata, Mohanpur 741246, Nadia District, West Bengal, India.

The surface states of a 3D topological insulator (TI) exhibit topological protection against backscattering. However, the contribution of bulk electrons to the transport data is an impediment to the topological protection of surface. We report the tuning of the chemical potential in the bulk of $Bi_2Se_2Te$ TI thin films, pinning it near the center of the bulk band gap, thereby suppressing the bulk carriers. The temperature dependent resistance of these films shows activated behavior down to 50K, followed by a metallic transition at lower temperatures, a hallmark of the robustness of TI surface states. Manifestation of topological protection and surface dominated transport is explained by 2D weak antilocalization phenomenon. We further explore the effect of surface to bulk coupling in TI in this work, which is captured by the number of effective conducting surface channels that participate in the transport. The presence of a single conducting channel indicates a strong surface to bulk coupling which is detrimental to purely topological transport. We demonstrate the decoupling of topological surface states on opposite surfaces of thin films, thereby suppressing the bulk transport. Our findings provide a deeper understanding of surface to bulk coupling along with topological transport behavior and their respective tunability.

## Introduction:

Topological Insulators (TIs) are a new class of quantum matter where the bulk is insulating and the surface consists of an odd number of Dirac cones which along with a Berry phase of $\pi$ on the Fermi surface enables them to exhibit numerous exotic phenomena such as topological protection against backscattering from non-magnetic impurities and defects that do not break time reversal invariance[1,2]. The Dirac like linear dispersion of these surface states (SS) along with helical spin texture arising out of spin-orbit coupling make them distinct from trivial surface states like 2D electron gas (2DEG) or other spin orbit coupled (SOC) systems. These states are robust to any amount of non-magnetic disorder and defects There are a number of evidences of surface dominated magneto-transport experiments[3–8] in thin films and nano-devices, arising out of two dimensional quantum interference phenomena like weak antilocalization (WAL), Shubnikov de – Hass (SdH) oscillations and Aharonov – Bohm (AB) Oscillations. These experiments provide direct evidence of topological protection of these robust surface Dirac states. WAL is a manifestation of quantum interference between electrons executing forward and time reversed paths, where the $\pi$ Berry phase causes a destructive interference in absence of magnetic impurities or external magnetic fields and prevents any localization brought about by backscattering of forward moving electrons into time reversed paths. However, the application of external magnetic field disrupts the time reversal symmetry resulting in localization of the electron. This phenomena is termed as weak antilocalization (WAL). WAL has been aptly described by Hikami, Larkin and Nagaoka (HLN) in two dimensional metals with spin-orbit interactions described later[9].

Second generation TI materials such as $Bi_2Se_3$ (BS), $Bi_2Te_3$ (BT) and $Sb_2Te_3$ (ST) called the binary TIs all suffered from parasitic bulk carriers which masks the contribution of surface states in conduction[10]. In the past few years there has been an intense search for better topological insulators than the earlier discovered binary TI materials, where often, there is significant bulk conduction owing to carriers generated by defects and Se/Te vacancies, which shifts the Fermi level in to the bulk[7,11]. Ternary and quaternary topological insulators materials, $Bi_2Se_2Te$ (BST), $Bi_2Te_2Se$ (BTS), $Bi_{2-x}Sb_xTe_3$, and $Bi_xSb_{1-x}Te_ySe_{1-y}$ (BSTS), on the other hand seem to be pushing the Fermi level inside the band gap[4,12–15]. ARPES and magnetotransport experiments validates the above point and hence these ternary tetradymites fulfill the criteria of intrinsic topological insulators[16–18]. Thus these materials are perfect candidates for realizing surface dominated transport leading to fabrication of spintronic devices harnessing the surface Dirac electrons alone. Currently the focus has shifted towards these materials in terms of compositional and band structure engineering owing to their superiority over their binary analogues in terms of yielding a stable insulating ground state[19].

The transport signature of the surface Dirac states have been achieved unambiguously with very less parasitic bulk effects in such ternary and quaternary TI flakes/single crystals[20–22]. Much of the transport studies on high quality BTS and BSTS has remained confined to single crystals or thin flakes exfoliated from these single crystals[4,5,14,18,20,23–25]. Although BTS and related Sb doped crystals have been found to be consisting of highly insulating bulk and highly mobile surface states, in some cases these crystals have been found to be metallic in nature as well, for example see ref. 11 & 29. Therefore, it is not certain that as grown crystals of BTS and BST will always show insulating character thus posing a serious uncertainty in growth parameters and compositions[11,25,26]. Both BST and BTS single crystals have been found to be insulating as well as metallic in nature with minor changes in composition ratio of Se/Te[13,14,18,27,28]. However, thin films of these TI's lack thorough characterization and investigation of the surface electron states by means of transport studies. To the best of our knowledge there are no studies on thin films of these bulk insulating alloys except for few where thin films of BTS deposited by Molecular Beam Epitaxy have been found to be metallic in nature[11,29] .

In this article we explore the transport properties of BST thin films grown by pulsed laser deposition (PLD) technique. Two insulating films of thickness 500nm (designated as B2) and 150nm (B1) and one metallic film (B3) of similar thickness to B1has been prepared for this study. The main findings of this paper are; fitting the resistance vs. temperature (R-T) curve with a parallel conduction mechanism, signatures of WAL and surface to bulk coupling from the study of R-T curve itself and the consistent description of MR data with the simplified Hikami, Larkin and Nagaoka (HLN) equation in the metallic as well as in insulating thin films[9]. The HLN equation (Eqn. 4) was formulated to capture WAL in 2D metals. This description removes the discrepancies concerning the values of transport coefficient "α" (see equation 4). "α" is a measure of the number of independent surface states contributing to the conductivity. Most papers report a single coherent channel and suggest that one of the surfaces has higher coherence length than the other. There have been efforts to control the number of channels by tuning the Gate voltage applied between the top and bottom surfaces[30]. Bulk carriers play a crucial role in TI coherent transport, and that channel coupling is controlled by a competition between the phase coherence time ($\tau_{\phi}$) and the surface-to-bulk coupling ($\tau_{SB}$). The no. of independent coherent channels can also be tuned by thickness, deposition conditions and temperature. PLD allows us to tune the Fermi level in the bulk band gap by varying these parameters, which is a cheaper analogue of the gating effect but sometimes more effective.

We fit the low field magneto-resistance (MR) data with the simplified HLN equation and the value of coefficient "α" remains around 0.5 for bulk metallic films, but changes from 0.5 to 1.0 in bulk insulating films corresponding to the two 2D conduction channels, the top and bottom surfaces, which are decoupled in the latter. With this finding we conclude that the surface states couple to the bulk in the films where the bulk is metallic, giving rise to one effective conduction channel, whereas, the bulk insulating films have no surface to bulk coupling and the two surfaces, top and bottom have separate contribution to the MR. Temperature dependence of α and phase coherence length ($l_{\varphi}$) for both the insulating as well as metallic thin films is also explained in this work.

## **Results and Discussions**:

We report the transport properties on three thin films of thicknesses 150nm (B1), 500nm (B2) and 150nm (C). The films "B1" and "B2" show insulating character while "B3", which is similar in thickness to B1, displays metallic character as shown in the inset of fig. 1(c). The film B3 was grown for the comparative study with the insulating films. The detailed film growth and characterization is described in supplementary.

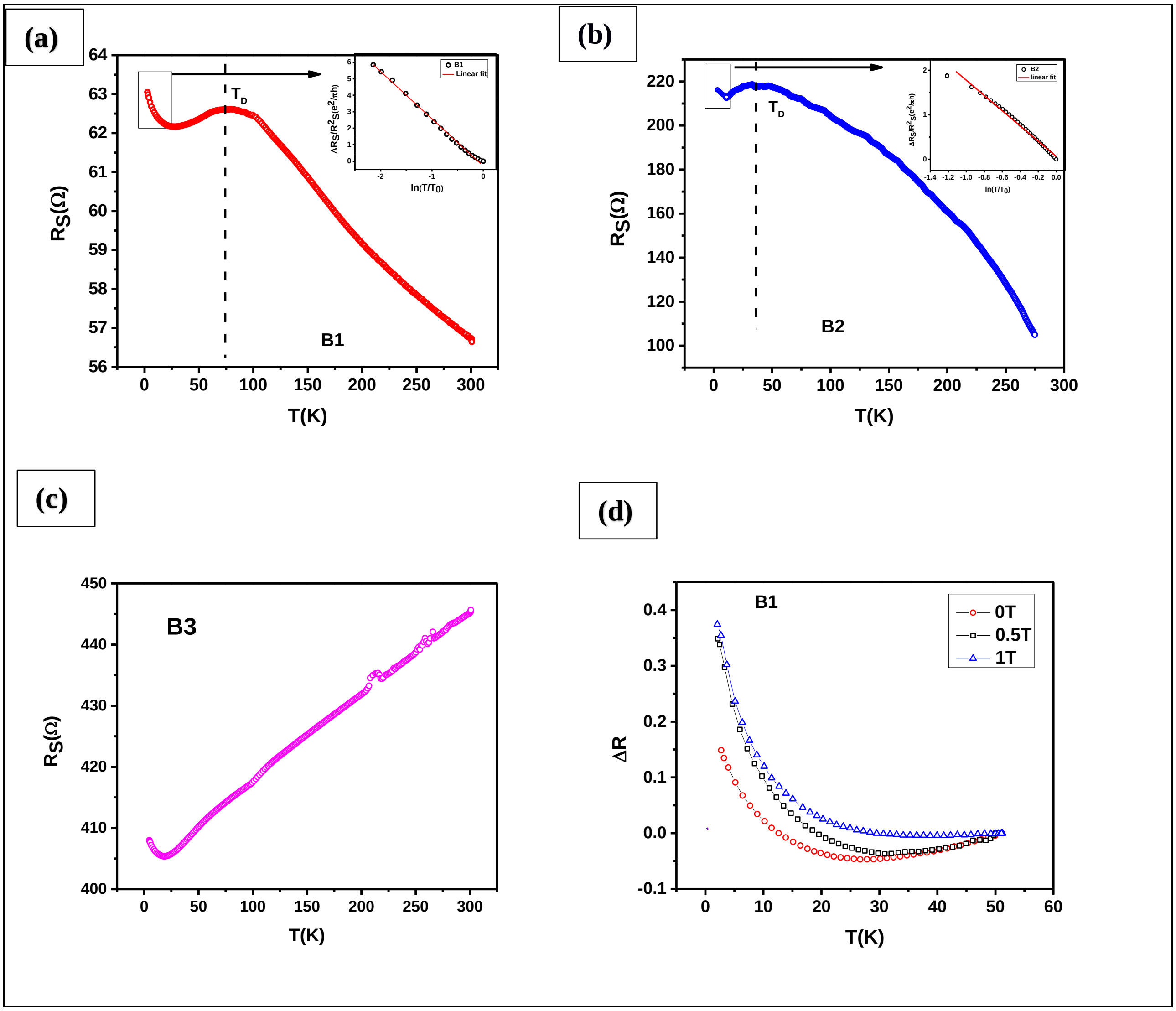

**Figure 1.** (a), (b) & (c) Resistance variation with temperature. This figure depicts the insulating and metallic behavior of BST thin films B1, B2 and B3 down to 2K. B1 and B2 display insulating character whereas B3 is metallic in nature (c). From the Hall measurements the carriers are found to be 'n' type (see supplementary section). The inset of B1 and B2 shows the logarithmic temperature dependent resistance upturn. (d) Application of perpendicular magnetic field changes the slope of the upturn in the R-T behavior for sample B1.

**Temperature dependent quantum correction:**

Interesting features of the electron transport can be inferred from the temperature dependence of resistance (Fig. **1**), especially in the low temperature regime. The existence of WAL depends on the competition between the phase coherence of the electron wave function and the dephasing factors present in the system. Temperature (electron- phonon scattering) and electron –electron scattering plays a crucial role in this dephasing mechanism at low temperatures. With lowering temperature as the phonon contribution to dephasing reduces in the system, a pronounced WAL type behavior in B1 and B2 can be seen from the resistance vs. temperature (R-T) data in fig1(a) and 1(b). This is captured from the portion below the dashed vertical line, which signifies the dominant role played by the surface states below this temperature. This can be considered as semiconductor to metal transition with decreasing temperature. Though traditionally, WAL is referred to in the context of MR data, it is also captured here in the R-T data below the dashed line as here the surface state dominated transport is protected from backscattering by the spin momentum locking and a π Berry phase.

In highly insulating TI samples WAL can be observed qualitatively by the nature of the R-T curve itself. This observation is somewhat analogous to the strong localization in two dimensional heterostructures where metal – insulator transition (MIT) can be captured by R-T alone[17,18,31,32]. One can draw an important conclusion from this observation that the fall in resistance in the R-T curve (below the indicated point $T_D$ on the graph) manifests robust delocalized nature of surface states against other interactions, disorders and grain boundaries at finite temperatures[33]. In our PLD grown BST thin films primary cause of disorder is from the granular structures (see SEM images in supplementary) which limit the mobility to a lower value of 25cm$^2$/V-sec. This in contrast to the single crystals and MBE deposited thin films which have mobilities of the order of 2000 – 4000cm$^2$/V-s, where surface Dirac electrons face nearly atomically flat surface. Therefore, PLD deposited thin films are real test for the topological protection of Dirac states against disorder.

Apart from downturn in $R_s$ (sheet resistance) from T= 70 and 50 K (denoted as $T_D$) in B1 and B2 thin films respectively, an upturn in the $R_s$ can also be observed at around 10K. This upturn in $R_s$ could be due to freezing of bulk carriers in the impurity band or e-e correlation among surface states on the surface of TI in presence of disorder or combined effect of these two mechanisms. In order to find out the possible mechanism behind this upturn one needs to separate the positive quantum correction due to WAL from the sheet conductance $G_S$. The WAL correction can be destroyed by application of perpendicular magnetic field as shown in the figure 1(d) for sample B1. The R-T curves in applied magnetic fields B =0.5 and 1.0 Tesla are distinct from the zero field R-T resulting in an upward shift, which is effected by the destruction of 2D WAL in the topologically protected surface states by the external field. The combined form of temperature dependent quantum corrections in the conductance due to WAL and e-e interaction can be written as [34,35];

$$\Delta G_S = \Delta G_{S(WAL)} + \Delta G_{S(e-e)} \tag{1}$$

$$\Delta G_S = \frac{\Delta R_S(T)}{R_S^2(T)} = c\frac{e^2}{2\pi^2\hbar}\ln(T) - \beta\frac{e^2}{2\pi^2\hbar}\ln(T) \tag{2}$$

$\Delta G_{S(WAL)} = c\frac{e^2}{2\pi^2\hbar}\ln(T)$ is a positive quantum correction due to 2D WAL effect[35]. Whereas e-e interaction driven negative correction in the conductance is given as; $\Delta G_{S(e-e)} = -\beta\frac{e^2}{2\pi^2\hbar}\ln(T)$. On fitting the temperature dependent correction in the conductance due to e-e interaction when WAL is destroyed (figure 1(d)), we get the value of the fitted interaction parameter β (see supplementary section). This value is not consistent with the theoretical value of the e-e interaction correction in the conductance in 2D [36]. It is therefore concluded that the upturn in these thin films is not entirely due to the e-e interaction but other mechanisms like localization or freezing out of the bulk electronic states[37].

The deviation of the coefficient β from the expected theoretically calculated values, indicates that even in the low temperature regime in the two insulating thin films (B1 and B2) do not show fully 2D transport and the total conduction is a combination of conduction of bulk and surface connected in parallel, henceforth referred to as the parallel channel conduction mechanism. The value of the parameter β could be used to estimate the truly 2D character of surface states of the bulk insulating TIs. Earlier electron-electron interactions have been observed in doped metallic thin films and ultrathin metallic films[35,38]. In these cases it could not be conclusively inferred whether the interaction is among surface states or the bulk states.

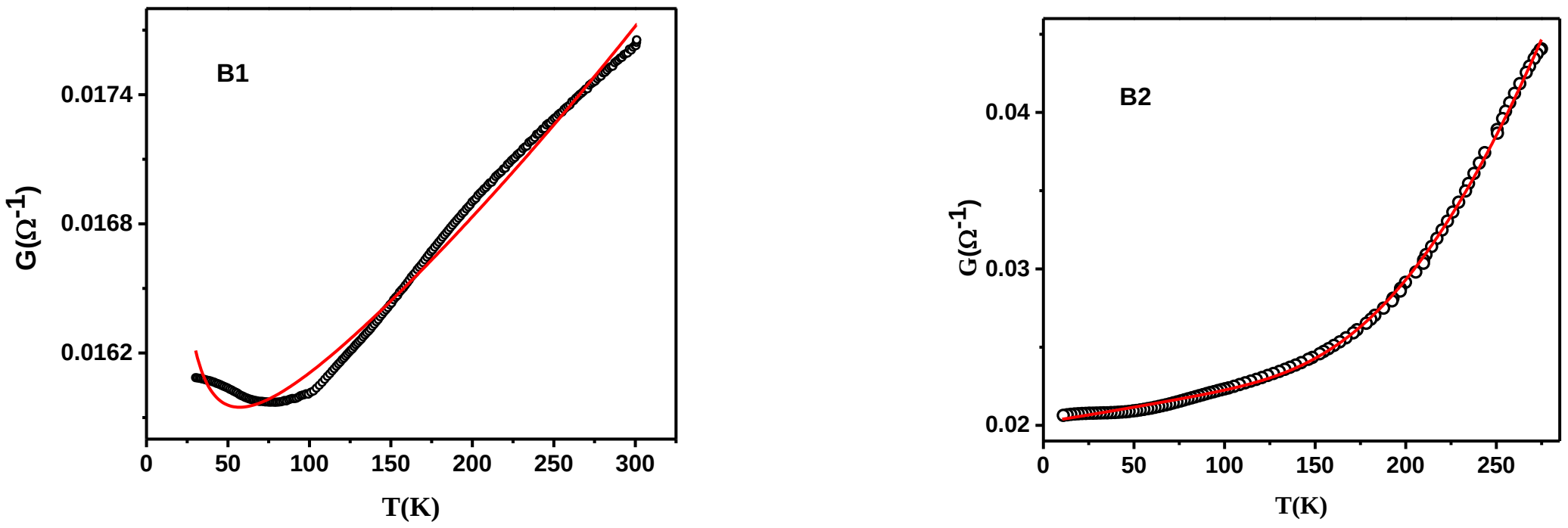

**Figure 2(a).** Conductance data fitted to a non-linear curve given in equation (3) for sample B2 which includes both surface and bulk contribution to the conductance. **(b)** Conductance as a function of temperature data fitted with the same expression. The deviation at the low temperature regime is due to the onset of WAL which yields a logarithmic correction to conductance (eqn. 2) which is not included in this fit.

Parallel channel conductance model has been used to fit the conductance vs T plot as shown in Fig. **2**. The two parallel channels being bulk conductance ($G_b$) and surface conductance ($G_S$). The combination of bulk and surface conductance, the total conductance; $G_t$ is represented by:

$$\begin{aligned} G_t &= G_b + G_S \\ &= 1/\left(R_S e^{\Delta/T}\right) + 1/\left(A+BT\right) \end{aligned} \quad (3)$$

$G_b(T) = 1/\left(R_S e^{\Delta/T}\right)$ *and* $G_S(T) = 1/\left(A+BT\right)$, where $\Delta$ represents the activation energy for impurity states in the bulk. The fitted temperature dependent conductance using equation (2) yields the value of $\Delta$ = 25.85 meV and $\Delta$ = 23.59 meV for samples B2 and B1 respectively. Coefficient A stands for disorder scattering which is independent of temperature, and B represents electron phonon scattering[46]. At high temperatures, conductance is dominated by the thermally activated carriers form the impurity band residing below the conduction band edge. As the temperature is reduced a freezing out of bulk carriers take place and surface conductance dominates the conduction in thin films.

Observation of various quantum interference phenomena in TI has been one of the important tools to characterize the backscattering immune topological surface states. In systems with large spin orbit interaction, having an odd number of Dirac cones the electron wave function gathers a “π” Berry phase as it circulates along a closed path on the iso-energic Fermi circles in a Dirac cone. This results in destructive interference between the two time-reversed coherent paths of electron waves in a diffusive medium, effectively increasing the probability amplitude for forward scattering. These quantum interference phenomena manifest itself in lowering the resistance or making positive quantum correction to the Drude conductivity at low temperatures in zero field. This is observed in the downturn in R-T plot below 70K and 50K in B1 and B2 respectively. This phenomenon is known as weak antilocalization as opposed to its counterpart weak localization.

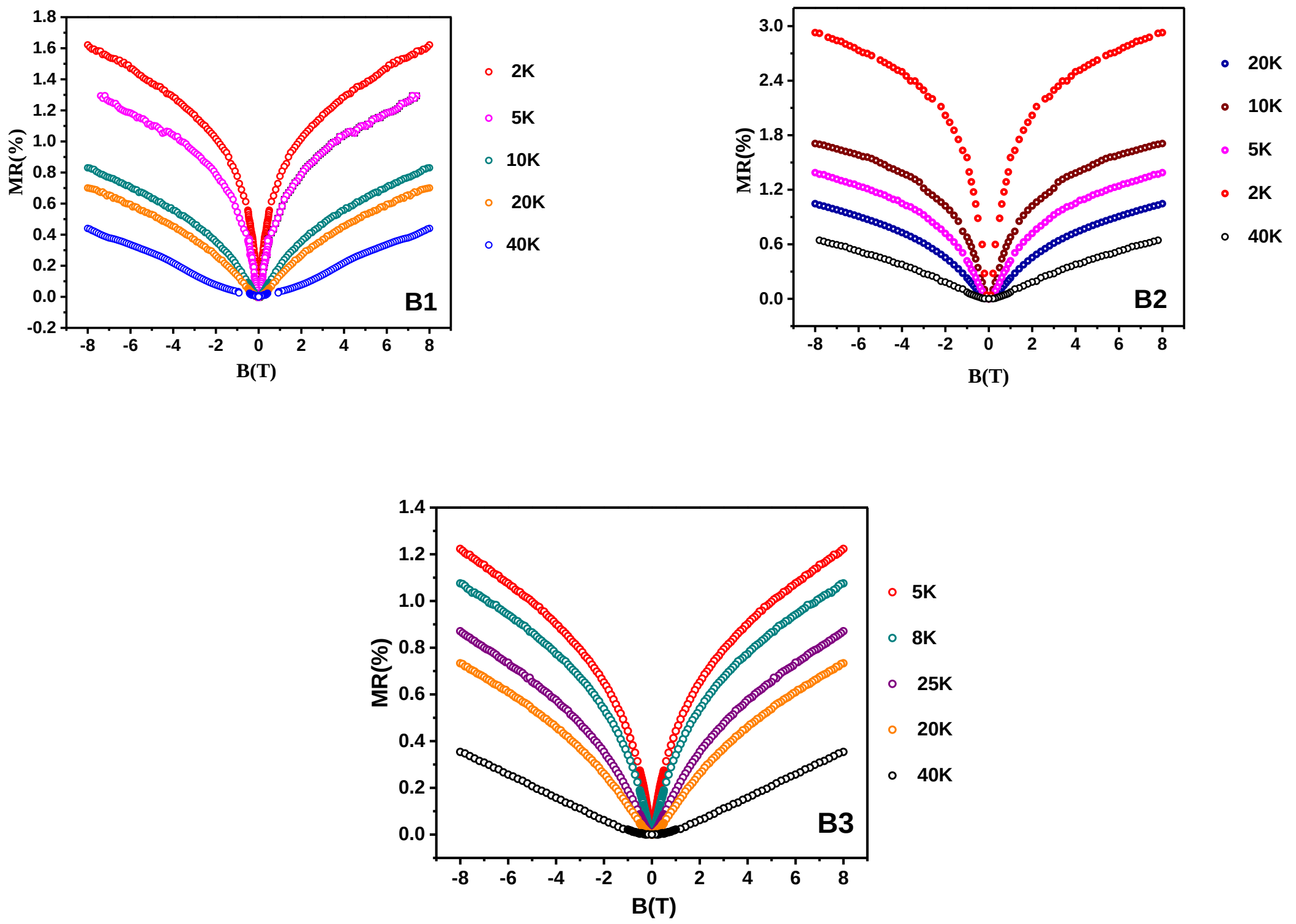

**Figure 3.** Magnetoresistance (MR) measured at different temperatures for the film B1, B2 and B3 thin films.  A sharp rise in MR around zero field (WAL) at lower temperatures is clearly seen in the above plot. B3 owing to metallic nature has the WAL cusp relatively less pronounced in comparison to B1 and B2.

**Magnetoresistance Data:**

From the temperature dependence of MR in fig. 3 it is clear that quantum coherent transport of surface electrons persists even at higher temperatures (20K), though reduced, showing a remarkable robustness of the surface states towards phonon scattering. The sharp dip like feature in MR at low temperature in all three samples is the manifestation of WAL and as the temperature increases this feature nearly vanishes above 40K resulting in parabolic MR which is classical in nature as opposed to WAL.

We first consider the low field (0 – 1T) correction to the conductance. In the presence of strong spin-orbit interaction, quantum correction to the conductivity in a two dimensional system can be described by HLN formula[39]. This expression is applicable in the diffusive quantum transport regime in samples with weak disorder, such as in thin films deposited by PLD. The HLN equation is valid for strong spin-orbit coupling limit and can be written in the simplified form for the following condition: $\tau_{so}<\tau_e<\tau_\varphi$, where $\tau_{so}$, $\tau_e$ and $\tau_\varphi$ are the characteristic time scales corresponding to the spin-orbit scattering, elastic scattering and inelastic scattering of charge carriers. Conductance correction due to WAL (HLN equation) consists of two components, the logarithmic and digamma components and is given as:

$$\Delta G_S = -\frac{\alpha e^2}{2\pi^2\hbar}\left[\Psi\left(\frac{B_\phi}{B}+\frac{1}{2}\right)-\ln\left(\frac{B_\phi}{B}\right)\right] \tag{4}$$

Where $\Psi$ is a digamma function and $B_\phi = \frac{\hbar}{4el_\varphi^2}$ is the characteristic magnetic field corresponding to the coherence length $l_\varphi$. The value of the coefficient "α" signifies effective number of independent 2D conducting channels and is related to the dimensionality of the conduction process in the system. For TI thin films the value of transport coefficient "α" should be equal to 1, a contribution of 0.5 corresponding to each of the two surfaces (the top and bottom surface) when they are decoupled. But experimentally this value has been found to vary from 0.3 to 1.5 and in some cases as large as 3.0[21]. The value of α=0.5 found in some earlier reports is similar to that found in thin films of Bismuth which is also a strongly spin orbit coupled semimetal but not a TI[40].

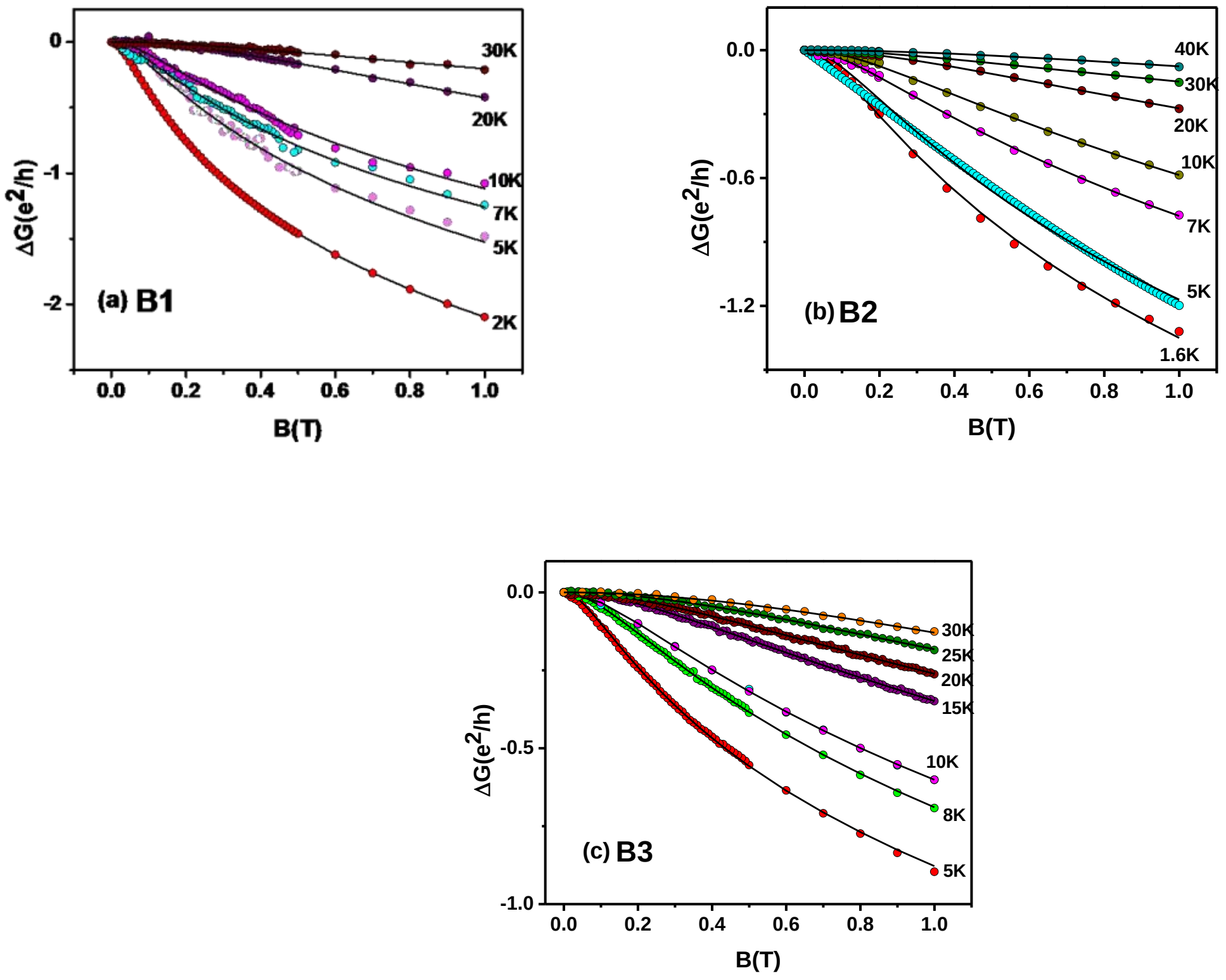

**<u>Figure 4.</u>** HLN fit of conductance data at different temperatures of samples B1, B2 and B3.

Fitting of magneto-conductance (MC) by HLN equation (4) in the low field regime at different temperatures is shown in the Fig. **4**(a), (b) and (c) for samples B1, B2 and B3 respectively and the corresponding extracted parameters "α" and phase coherence length "$l_\phi$" are plotted as a function of temperature in Fig. **5**(a), (b) and (c). In the bulk insulating and decoupled regime we have observed strong robustness of the conduction channel parameter "α" which hovers around "1.0" consistent with the two surface transport. This points towards a two independent channel coherent transport. Whereas in metallic films it remains below 1 and "α" around "0.7". This value is consistent with ultrathin TI films where the top and bottom surfaces are directly coupled and in thick TI films with metallic bulk (mainly BS and BT), where the coupling is mediated by the bulk electrons[41]. This is in contrast with the earlier case, where the top and bottom surfaces are decoupled with $\alpha = 1$. It should be noted here, WAL characteristics in these films can only be seen at low temperatures as evident from the sharp dip in MC at low field. At higher temperatures it takes parabolic nature (a broaden cusp) where surface and bulk both contribute to MC. Therefore fitting the MC curves with simple HLN model at higher temperature is not appropriate. Therefore in order to capture the WAL from surface states at relatively higher temperatures we fit MC data with modified HLN equation with a parabolic term (see supplementary).

Earlier studies based on the two channel conduction in TI thin films and gated devices in ultrathin and thick limit have shown that the value of the coefficient "α" depends on various parameters such as gating voltage, compensation doping, and thickness of the film[30,34,42–44]. At sufficiently low temperature i.e. 2K, we can expect that the bulk electrons will not contribute to the conductivity (in B1 and B2) as evident from the thermal activation energy of the order of $\Delta$ = 25.85 meV, where there are hardly any excitations from the impurity band to the bulk conduction band. This energy scale is much higher for any excitation at the lowest temperatures and there is no bulk electronic conduction at these temperatures.

Variation of the coefficient α with temperature is shown in fig. 5. For the bulk insulating films, B1 and B2, we see that α decreases from 1 to 0.5, whereas of the bulk metallic film B3 it increases. In the bulk insulating case, at very low temperatures the number of bulk carriers is low (see Hall data in supplementary) and the top and bottom surface channels are nicely decoupled from the bulk, giving rise to the value of α to '1'. From the perspective of WAL, the surface and bulk states can be regarded as independent phase coherent channels as long as the carriers in one channel lose coherence before being scattered into the other. Here, the majority of closed loops responsible for WAL will involve states from a single channel, and each channel will exhibit its own WAL correction. This condition can be formulated as $\tau_{SB} > \tau_{\varphi}$, where $\tau_{SB}$ is the effective surface-to-bulk scattering time and $\tau_{\varphi}$ the coherent dephasing time. In the opposite limit $\tau_{SB} < \tau_{\varphi}$, charge carriers scatter between the bulk and surface states while maintaining phase coherence, effectively becoming a single phase-coherent channel. Thus the temperature dependence of α is related to the decrease in $\tau_{SB}$, brought about by the increase in number of carriers in the bulk with temperature owing to thermal excitations. This results in increase in surface to bulk coupling and the separation between the top and bottom surface is gradually diminished with an increase in the temperature effectively increasing the effective conduction channels. Once we are at a fairly high enough temperature, the top and bottom surfaces and the bulk behave as a single coherently coupled conduction channel and hence the effective α is '0.5'. The coefficient α can be used phenomenologically as a measure for channel separation. α decreases with temperature, suggesting a decreasing channel separation with temperature. The effective number of phase coherent channels reflected in the magnitude of α depends on the ratio $\tau_{SB}$ to $\tau_{\varphi}$, which was controlled through changing the bulk carrier density (as reflected in our Hall data reported in the supplementary), which plays an important role in the surface to bulk coupling. An intuitive picture of surface to bulk coupling is given by in fig. 3(d) of ref. 31. Steinberg *et. al.*, have varied the gate voltage to tune the $\tau_{\varphi}/\tau_{sb}$ ratio capturing a somewhat similar variation in "α", whereas we were able to achieve a similar tuning by varying the temperature and deposition parameters[30].

In the bulk metallic case (B3) on the other hand, there are enough bulk carriers even at the lowest temperature and the system behaves effectively as a single coherently coupled conducting channel reflected in an α of '0.5'. In this case the $\tau_{SB} < \tau_{\varphi}$, one can see that the carrier density obtained from Hall data (suppl) is independent of temperature. Hence one can surmise that the $\tau_{SB}$ is temperature independent. However, as $l_{\varphi}$ goes down with temperature (Fig. 6), so does $\tau_{\varphi}$ ($l_{\varphi} = \sqrt{D\tau_{\varphi}}$), the $\tau_{\varphi}/\tau_{SB}$ ratio increases with temperature and the magnitude of α goes up to 1.

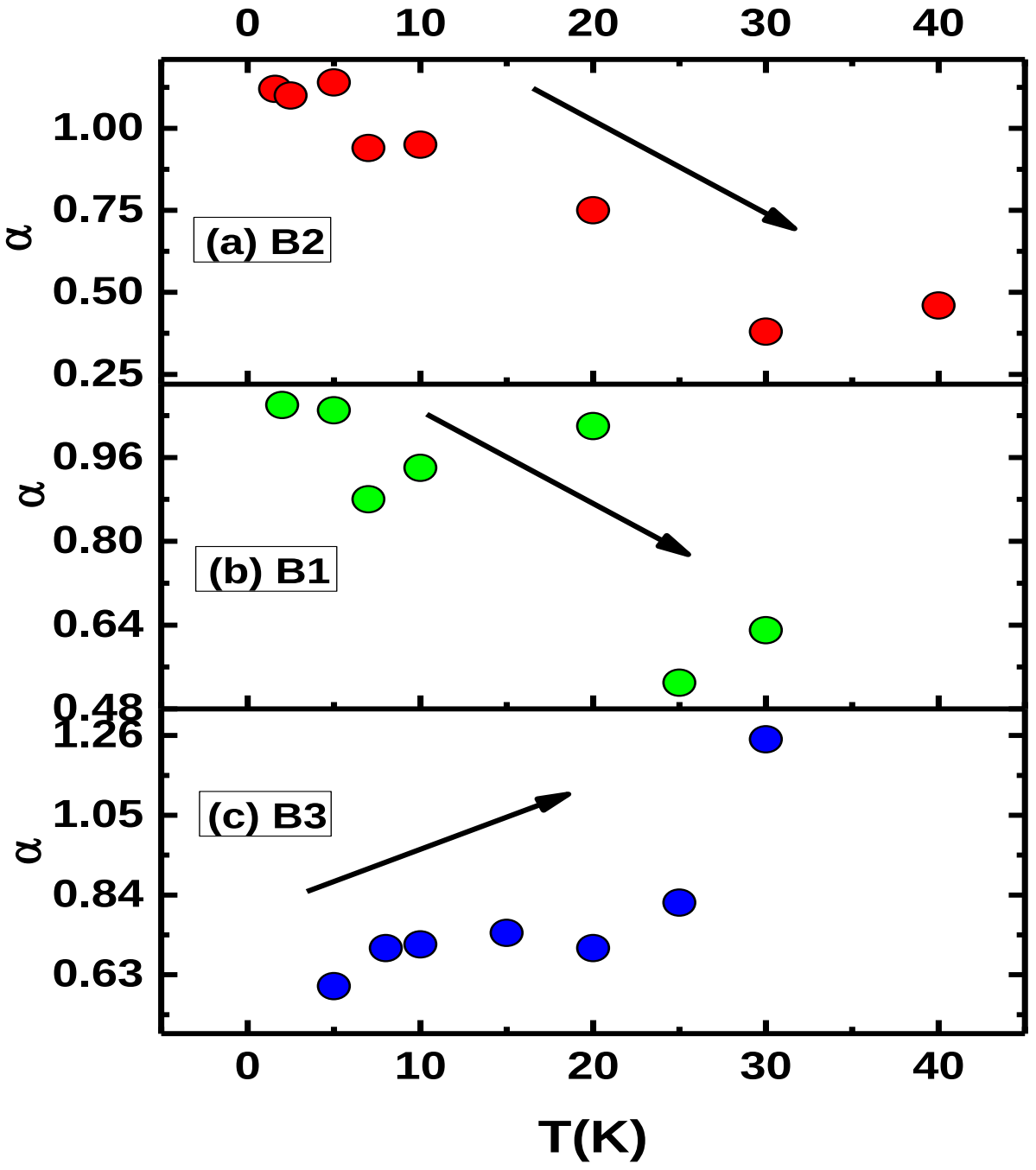

**Figure 5.** Variation of α with temperature for the three films B2, B1 and B3. For the insulating samples the behavior of α is opposite to that for the metallic film B3. The direction of the arrow signifies increasing or decreasing trend of α for metallic or insulating films.

Temperature dependent variation of phase coherence length ($l_\varphi$) is shown in fig. 6 for all the three samples. While the temperature variation of "$\boldsymbol{\alpha}$" in B1/B2 and B3 is different, $l_\varphi$ decreases monotonically with increasing temperature (fig. 6) in all the films. This is a manifestation of dephasing by phonons on the surface of the film. From the SEM images taken of these films, the average size of the grains on the thin films is larger than the coherence length $l_\varphi$, which implies that granular structures of the films play a minimal role in dephasing surface electrons. The variation of $l_\varphi$ with temperature signifies 2D surface dominant transport. Phase coherence length decreases with temperature with a power law variation $l_\varphi \sim T^{-p}$, where p depends on the dimensionality of the system and the agents responsible for the dephasing mechanism[45,46]. For the three films under study in our case i.e., B1, B2 and B3 the value of p = 0.61, 0.31and 0.64 respectively[37]. Detailed investigation of this coefficient 'p' is being carried out for a variety of films. This will be reported elsewhere.

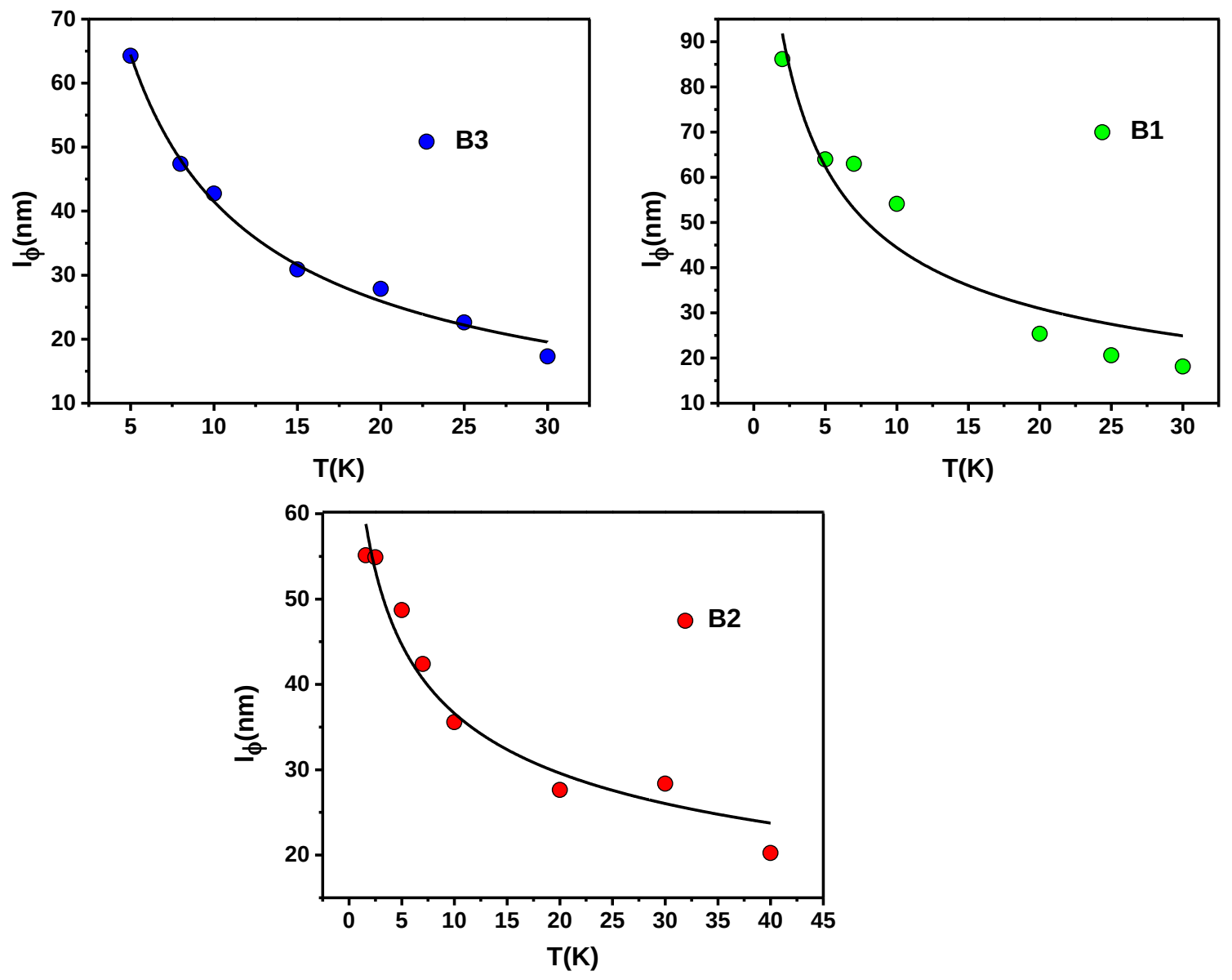

**Figure 6.** Phase coherence length vs. temperature for the three thin films B3, B2 and B1. The variation with temperature follows the expected power law behavior with $l_\varphi \sim T^{-p}$.

We have also fitted the magneto-conductance (MC) curves up to high magnetic fields (8T) with the modified HLN equation where we added the classical Kohler term ($MC \propto B^2$) to it. This along with fitting to higher temperature MC curves (40 K and 100K) are shown in the supplementary section. In the low temperature MC, at high fields the mixture of the classical Kohler term and the quantum MC gives the MC a linear appearance. At intermediate temperatures (40K) the classical nature kicks in at lower fields in comparison to the low temperature curves and hence the classical MC dominates giving it a somewhat parabolic appearance (Fig. 3). The exact reason for the linear MC in the high field regime is still debated [47] and will be addressed by us in details elsewhere. Though some reports attribute linear nature of MC to the surface states but at such a high temperatures (say 100 K), there can be sizable bulk carrier contribution and electron-phonon scattering which may override the surface Dirac state contribution to MC. Therefore we can attribute it to the mixed phase of MR arising from classical as well as quantum coherence effects.

## Conclusion:

It is remarkable that PLD grown films, studied in this work are insulating in the bulk and exhibit topological transport behavior, clearly manifested in the bulk insulating nature of R vs. T curves. PLD is a better deposition technique in comparison to other synthesis techniques as it is cost effective and offers ready integration with other materials for device applications. Our films were prepared to be rather thick (500 nm – 150nm) to facilitate a direct decoupling of the two surface states with the bulk such that the condition $\tau_\varphi / \tau_{sb} < 1$ is satisfied at low temperatures, where the value of the transport channel coefficient "α" was found to be 1 for these decoupled samples. We were able to tune the $\tau_\varphi / \tau_{sb}$ ratio by varying the temperature, thereby tuning "α". This was studied for both bulk insulating and bulk metallic films in contrasting ways. A coherent and comprehensive understanding has been given for the different values of the transport coefficient "α" for different experimental conditions (temperature variation) and sample properties, i.e., for both bulk insulating and metallic films. This gives a better understanding of the underlying physics of topological insulator which coupled with tunability of PLD can lead to synthesis of better TI materials and devices.

## Methods:

Thin films of BST were grown by Pulse Laser Deposition method, where BST targets were ablated using high power Excimer Laser and deposited on Si (100) substrate. The thickness of the samples was controlled by monitoring the number of pulse shots on the target and the thickness was verified by profilometer. The devices were fabricated using optical lithography technique and a standard Hall bar pattern was inscribed on top of the samples. Indium contacts were put on the sample and then the samples were subjected to transport measurements down to low temperatures (300K-1.6K) and under high magnetic field (8T).

**Acknowledgements**

Authors would like to acknowledge IISER Kolkata and Ministry of Human Resource Development for financial support for this experiment. SS and JS would like to thank University Grants Commission for fellowship and contingency grants.

**Author Contributions**

R.K.G prepared the samples and performed the experiments. S.S and J.S. analyzed the data. A.M. did the lithography and contacts. C.M. conceived and supervised the whole work and wrote the manuscript.

**Additional Information**

Competing financial interests: The authors declare no competing financial interests.